
\documentstyle{amsppt}
\magnification=1200
\hsize=5.5truein
\vsize=8.15truein
\voffset 0.25truein
\TagsOnRight
\baselineskip=14pt
\parindent 1.5em
\nopagenumbers

\document

\rightline {McGill/94-13}

\rightline {hep-th/9402132}

\bigskip

\bigskip

\bigskip

\centerline {\bf GRADED CONTRACTIONS OF BILINEAR INVARIANT\/}

\centerline {\bf FORMS OF LIE ALGEBRAS\/}

\bigskip
\bigskip
\centerline {Marc de Montigny}

\centerline {Department of Physics}

\centerline {McGill University}

\centerline {Montreal (Quebec)}

\centerline {H3A 2T8, CANADA}

\bigskip
\bigskip
\bigskip
\centerline {\bf February 1994\/}
\bigskip
\bigskip
\bigskip

ABSTRACT.
 We introduce a new construction of bilinear invariant forms on
 Lie algebras, based on the method of graded contractions.
 The general method is described and the $\Bbb Z_2$-,
 $\Bbb Z_3$-, and $\Bbb Z_2\otimes\Bbb Z_2$-contractions are found.
 The results can be applied to all Lie algebras and superalgebras
 (finite or infinite dimensional) which admit the chosen gradings.
 We consider some examples: contractions of the Killing form,
 toroidal contractions of $su(3)$, and we briefly discuss
 the limit to new WZW actions.

\bigskip

\leftline {PACS numbers: 02.20.+b, 11.30.--j}

\bigskip

\bigskip

\leftline {Submitted to {\it J. Phys. A: Math. Gen.}}

\newpage

\leftline {\bf 1. INTRODUCTION.}
\medskip

Contractions of Lie algebras were introduced in physics
 forty years ago by Wigner and In\"on\"u in order to
 get a formal way of passing from the Poincar\'e
 group to the Galilei group \cite{1}.
 In general, contractions consist of the introduction
 of parameters in the basis of
 a Lie algebra such that, for some singular value of these
 parameters, we get a different ({\it i.e.} nonisomorphic) algebra.
 The interest of this method
 in physics stems from the fact that it relates
 different symmetry groups, from which the {\it contracted}
 group can be understood as an ``approximation'' of the
 ``exact'', or noncontracted group. (A discussion of the concept
 of contraction and
 some generalizations of the Wigner-In\"on\"u method are
 given in Ref. \cite{2}).

A different approach has been developed recently, which is based
 on the preservation of some {\it grading} of the Lie algebra through
 the contraction procedure. The general method of so-called
 {\it graded contractions} is presented in
 Ref. \cite{3}. The analogous theory for the representations
 (which contains the contractions of algebras as a particular case)
 and their Casimir operators is found in Refs \cite{4} and
 \cite{5}, respectively.

In this paper we use the concept of graded contractions of Lie
 algebras to construct new symmetric bilinear invariant forms of a Lie
 algebra (in general, non-semisimple), starting from the known
 bilinear form of a noncontracted algebra. Even within a single
 Lie algebra, one can obtain different bilinear forms starting
 from one bilinear invariant form.
 As for the method of
 graded contractions of algebras, the present procedure is
 very general and can be applied to any
 Lie algebra or superalgebra, of finite or
 infinite dimension.
 Also the results are universal in the sense that
 given a grading group
 (which is abelian for our purpose), the problem is solved simultaneously
 for all Lie algebras of any type and dimension
 which admit the chosen grading.
 The construction is presented in the next section. In Section 3
 we find the general bilinear invariant form obtained from preserving
 $\Bbb Z_2$-, $\Bbb Z_3$-, and $\Bbb Z_2\otimes\Bbb Z_2$-gradings.
 In Section 4 we present some examples: contractions of the Killing
 form, toroidal contractions of $su(3)$ (or, more precisely, of
 its complexification $A_2$), and the ``deformation'' of
 Wess-Zumino-Witten (WZW) actions (defined over some group manifold)
 into modified WZW actions, defined over a contracted group.

\medskip

\leftline {\bf 2. DESCRIPTION OF THE METHOD.}
\medskip

 Consider a Lie  algebra $g$  (corresponding to a Lie group $G$)
 defined over the field $\Bbb K \ (=\Bbb R$ or $\Bbb C\ )$.
 A symmetric bilinear invariant form $\Omega$ on $g$ is a mapping,
$$
\Omega : \quad g\times g\rightarrow \Bbb K\ ,\tag2.1$$
which satisfies the symmetry property,
$$
\Omega (X, Y) = \Omega (Y, X),\tag2.2$$
and is $G$-invariant,
$$
\Omega (\vartheta X\vartheta^{-1},\vartheta Y\vartheta^{-1}) =
 \Omega (X, Y),\tag2.3$$
with $X, Y\in g$, and $\vartheta\in G$.

In terms of the generator $Z$
 of $\vartheta$ in the Lie algebra, (2.3) is equivalent to
$$
\Omega (X, [Z, Y]) + \Omega (Y, [Z, X]) = 0.\tag2.4$$

Given a basis of generators $\{X_1,\dots ,X_{dim[g]}\}$ of $g$ such
 that $[X_A, X_B] = f^C_{AB} X_C$, the
 bilinear form is often written as a matrix $\Omega_{AB}\equiv
 \Omega (X_A, X_B)$ which is symmetric (from (2.2)) and
 satisfies $f^D_{AB}\Omega_{CD} + f^D_{AC}\Omega_{BD} = 0$
 (from (2.4)).

Now let us recall some definitions and properties that we need from
 the theory of graded contractions of Lie algebras (more details are
 in Refs \cite{3, 4}). A
 $\Gamma$-grading of a Lie algebra $g$ consists of the
 decomposition
 $$
g=\bigoplus_{\mu\in\Gamma} g_\mu,\tag2.5$$
into eigenspaces under the action of a set of automorphisms of finite
 order on $g$ (for our purposes we can take $\Gamma$ to be an abelian
 finite group).
 Each of these automorphisms provides $g$ with a grading
 by a cyclic group having the same order. Thus the grading group $\Gamma$
 is the tensor product of all the cyclic groups associated with every
 automorphism: $\Gamma = \bigotimes \Bbb Z_N$.

The commutators in $g$ inherit a grading structure from the
 automorphisms of finite order, that is, if $X\in g_\mu$ and
 $Y\in g_\nu$, then
$$
[X, Y] = Z,\tag2.6$$
 with $Z\in g_{\mu +\nu}$ as long as the commutator is not zero.
 (The addition $\mu +\nu$ denotes the product of elements
 $\mu$ and $\nu$ in the grading group
 $\Gamma$.)
 This grading structure can be written symbolically,
$$
0\neq [g_\mu , g_\nu ]\subseteq g_{\mu + \nu },
\qquad \mu ,\ \nu ,\ \mu +\nu\in\Gamma.\tag2.7$$

A graded contraction $g^\varepsilon$ of $g$ is defined by modifying
 the commutators,
$$
[g_\mu ,g_\nu ]_\varepsilon\equiv\varepsilon_{\mu ,\nu}
 [g_\mu ,g_\nu ]\subseteq \varepsilon_{\mu ,\nu}g_{\mu +\nu},
\tag2.8$$
where $\varepsilon\in\Bbb K$ . We
 have kept the $\varepsilon$-parameter in the
 right-hand side of
 (2.8) in order to show that if $\varepsilon =0$, then
 the commutator of an element of $g_\mu$ with an element of $g_\nu$
 vanishes.
 From the definition (2.8) we see
 that the $\varepsilon$ parameters are symmetric
 ({\it i.e.} $\varepsilon_{\mu ,\nu} = \varepsilon_{\nu ,\mu}$), and
 by enforcing the Jacobi identities on the new commutators, one finds
 that the $\varepsilon$-parameters must satisfy the {\it contraction
 relations},
$$
\varepsilon_{\mu ,\nu}\varepsilon_{\mu +\nu , \lambda} =
\varepsilon_{\nu ,\lambda}\varepsilon_{\nu +\lambda ,\mu},\tag2.9$$
for all the values of the indices. A solution $\varepsilon$ of
 (2.9) defines a contraction, that is, a new Lie algebra.

Now we define the contractions of the bilinear invariant form $\Omega$.
 Using the same notation as in
 (2.7), we define the {\it contracted symmetric
 bilinear invariant form} $\Omega^\gamma$ as
$$
\Omega^\gamma (g_\mu , g_\nu ) \equiv
 \gamma_{\mu , \nu}\Omega (g_\mu , g_\nu ),\tag2.10$$
where $g_\mu \ (g_\nu )$ represents any element $X\in g_\mu
 \ (\in g_\nu)$.

{}From the properties of symmetry and invariance, and given a
 contraction matrix $\varepsilon$
 ({\it i.e.} a contracted algebra), we get the restrictions upon
 $\gamma$:
$$
\gamma_{\mu ,\nu} = \gamma_{\nu ,\mu}\tag2.11$$
and
$$
\varepsilon_{\lambda ,\mu}\gamma_{\lambda +\mu ,\nu} =
\varepsilon_{\lambda ,\nu}\gamma_{\lambda +\nu ,\mu}.\tag2.12$$

\medskip

\leftline {\it Proof.}

The relation (2.11) follows from equations (2.2) and (2.10).

The condition (2.12) is obtained from substituting (2.10) into (2.4):

$$
\aligned
&\Omega^\gamma (g_\mu ,[g_\lambda ,g_\nu ]_\varepsilon )
 + \Omega^\gamma (g_\nu ,[g_\lambda ,g_\mu ]_\varepsilon ) = 0,\\
&\Omega^\gamma (g_\mu ,\varepsilon_{\lambda ,\nu}[g_\lambda ,
 g_\nu]) +
 \Omega^\gamma (g_\nu ,\varepsilon_{\lambda ,\mu}[g_\lambda ,
 g_\mu])  = 0,\\
&\varepsilon_{\lambda ,\nu}\gamma_{\mu ,\lambda +\nu}
 \Omega (g_\mu ,[g_\lambda ,g_\nu ]) +
 \varepsilon_{\lambda ,\mu}\gamma_{\nu ,\lambda +\mu}
 \Omega (g_\nu ,[g_\lambda ,g_\mu ]) = 0.
\endaligned $$
 By comparing the last line with the invariance condition relation,
 $$\Omega (g_\mu ,[g_\lambda ,g_\nu ]) + \Omega
 (g_\nu ,[g_\lambda ,g_\mu ])=0,$$ before contraction, we see that one
 must have $\varepsilon_{\lambda ,\nu}\gamma_{\mu ,\lambda +\nu}
 = \varepsilon_{\lambda ,\mu}\gamma_{\nu ,\lambda +\mu}$. The
 form (2.12) is obtained by using the symmetry properties of
 $\varepsilon$ and $\gamma.\ \ \bullet$

\medskip

The solutions of (2.11, 12) provide new invariant bilinear forms,
 obtained by substituting the $\gamma$-parameters back into (2.10).
 As for the contractions of algebras and representations, there
 is a trivial contracted bilinear form, for which all $\gamma$'s
 are equal to zero. However, the solution with all
 the $\gamma$'s equal to one is not trivial. As explained in
 Ref. \cite{3}, the equations in (2.9) which contains an undefined
 $\varepsilon$ (when the corresponding commutator vanishes
 already in the noncontracted algebra) does not appear in the
 system of equations to be solved. The same occurs for the $\gamma$'s.
 If $\mu$ and $\nu$ are such that $\Omega (g_\mu , g_\nu) =0$, then
 the equations containing the corresponding $\gamma_{\mu ,\nu}$ must be
 removed from the system (2.12). If it does not happen, we say that
 this grading is {\it generic} regarding the
 invariant bilinear form.  Note
 that definitions (2.8) and (2.10) allow one to consider also
 {\it deformations}, that is, processes after which some commutators
  or bilinear forms
 --initially zero-- become non-trivial, but satisfy the grading
 property nevertheless. We shall not consider this type of process
 below.

Note also that the ``composition'' $(\gamma_1\bullet\gamma_2)_{\mu ,\nu}
 \equiv (\gamma_1)_{\mu ,\nu} (\gamma_2)_{\mu ,\nu}$ (without summation
 over repeated indices) of solution matrices, which we have defined for
 the contractions of algebras in equations (2.8, 9) of Ref. \cite {3}, here
 does not yield a new
 solution in general. Thus, given two solutions of (2.12),
 $\gamma_1$ and $\gamma_2$, their composition will not be a solution
 unless extremely particular conditions on
 the $\varepsilon$'s are satisfied.
 The same applies for the ``normalization'' of the $\gamma$'s, because
 it is tied up to the normalization of the $\varepsilon$'s. If we
 perform a change of basis $g_\mu\rightarrow g'_\mu = a_\mu g_\mu$, then
 the invariant bilinear form becomes $\Omega^\gamma (g'_\mu ,g'_\nu )
 = {\gamma_{\mu ,\nu}\over {a_\mu a_\nu}}\Omega (g_\mu ,g_\nu )$.
 Unless we consider very particular values of $\gamma$'s and
 $\varepsilon$'s it will not be possible to satisfy ${\gamma_{\mu ,\nu}
 \over {a_\mu a_\nu}} = 1$ for all $\gamma$'s. Unlike the contractions
 of representations, in which case we can normalize the parameters
 $\psi$ (see Ref. \cite {4}) by changing the basis of the
 representation vector space independently
 of the basis of the algebra, here the
 normalizations of $\varepsilon$ and $\gamma$ are interdependent.

\medskip

\leftline {\bf 3. $\Bbb Z_2$-, $\Bbb Z_3$-, AND
 $\Bbb Z_2\otimes\Bbb Z_2$-CONTRACTIONS.}
\medskip

In this section we find the contractions of the invariant bilinear forms,
 using the results of Ref. \cite{3}, for Lie algebras over the
 field of {\it complex} numbers.
 As in Ref. \cite{3} we consider the contractions with gradings
 of {\it generic} type. We have not normalized the $\gamma$'s in order
 to display the possible zero parameters. (Note that the
 $\varepsilon$'s of the present
 paper are the $\gamma$'s in Ref. \cite{3}.)

\medskip

\leftline {\bf 3.1. $\Bbb Z_2$-contractions.}
\medskip

The grading group consists of two elements,
 $\Gamma = \{0,1\}$, with the product (denoted additively):
 $$0+0=0,\quad 0+1=1+0=1,\quad 1+1=0.$$ There are three independent
 $\gamma$-parameters: $\gamma_{0,0}, \gamma_{0,1}\ (=\gamma_{1,0})$
 and $\gamma_{1,1}$. We shall cast them into a matrix form:
 $\gamma =\pmatrix
 \gamma_{0,0} & \gamma_{0,1}\\
 \gamma_{0,1}  & \gamma_{1,1}\endpmatrix$, although it should not be taken
 formally as a matrix, since none of the properties of matrices
 --multiplication, inverse, determinant, etc.-- are shared by the
 present objects.
 (Hereafter, we write only the upper
 diagonal of $\gamma$ and $\varepsilon$,
 remembering that they are symmetric).

The condition (2.12) reads, for $\Bbb Z_2$,
$$
 \varepsilon_{0,0}\gamma_{0,1} = \varepsilon_{0,1}\gamma_{0,1},\quad
 \varepsilon_{0,1}\gamma_{1,1} = \varepsilon_{1,1}\gamma_{0,0}.
\tag3.1$$

For the trivial contraction $\varepsilon =\pmatrix
 \varepsilon_{0,0} & \varepsilon_{0,1} \\
   & \varepsilon_{1,1}\endpmatrix = \pmatrix
 1 & 1 \\
   & 1 \endpmatrix = (1)$, the relations (3.1) become $\gamma_{0,1} =
 \gamma_{0,1}$ and $\gamma_{0,0}=\gamma_{1,1}$
 (which imposes no further restriction upon
 $\gamma_{0,1}$).
For the other trivial contraction $\varepsilon = \pmatrix
 0 & 0 \\
   & 0\endpmatrix =(0)$, there are no restrictions on $\gamma$, which
 are thus completely free.
If $\varepsilon_{0,0}\neq\varepsilon_{0,1}$,
 then $\gamma_{0,1}$ must vanish. This happens for the
 Wigner-In\"on\"u-like contraction matrix
 $\varepsilon = \pmatrix
 1 & 0 \\
   & 0\endpmatrix$.

In summary, we find the following contractions,
$$
\aligned
\varepsilon &= \pmatrix
 1 & 1 \\
   & 1 \endpmatrix ,\
\gamma = \pmatrix
 a & b \\
   & a \endpmatrix ,\\
\varepsilon &= \pmatrix
 0 & 0 \\
   & 0 \endpmatrix ,\
\gamma = \pmatrix
 a & b \\
   & c \endpmatrix ,\\
\varepsilon &= \pmatrix
 1 & 0 \\
   & 0 \endpmatrix ,\
\gamma = \pmatrix
 a & 0 \\
   & b \endpmatrix ,\\
\varepsilon &= \pmatrix
 0 & 0 \\
   & 1 \endpmatrix ,\
\gamma = \pmatrix
 0 & a \\
   & b \endpmatrix ,\\
\varepsilon &= \pmatrix
 1 & 1 \\
   & 0 \endpmatrix ,\
\gamma = \pmatrix
 a & b \\
   & 0 \endpmatrix .\endaligned\tag3.2$$
where $a, b$ and $c$ are arbitrary (possibly zero) complex numbers.

\medskip

\leftline {\bf 3.2. $\Bbb Z_3$-contractions.}
\medskip

The $\Bbb Z_3$-contractions ($\Bbb Z_3$ consists of
 three elements $0, 1, 2$ on which the product is the usual addition
 modulo 3) are determined by six parameters (cast into
 matrix form),
$$
\gamma =\pmatrix
 \gamma_{0,0} & \gamma_{0,1} & \gamma_{0,2} \\
   & \gamma_{1,1} & \gamma_{1,2} \\
   &   & \gamma_{2,2} \endpmatrix .$$

In terms of these parameters, the restrictions (2.12) are
$$
\aligned
&\varepsilon_{0,0}\gamma_{0,1} = \varepsilon_{0,1}\gamma_{0,1},\quad
\varepsilon_{1,1}\gamma_{2,2} = \varepsilon_{1,2}\gamma_{0,1},\\
&\varepsilon_{0,0}\gamma_{0,2} = \varepsilon_{0,2}\gamma_{0,2},\quad
\varepsilon_{0,2}\gamma_{1,2} = \varepsilon_{1,2}\gamma_{0,0},\\
&\varepsilon_{0,1}\gamma_{1,2} = \varepsilon_{0,2}\gamma_{1,2},\quad
\varepsilon_{0,2}\gamma_{2,2} = \varepsilon_{2,2}\gamma_{0,1},\\
&\varepsilon_{0,1}\gamma_{1,1} = \varepsilon_{1,1}\gamma_{0,2},\quad
\varepsilon_{1,2}\gamma_{0,2} = \varepsilon_{2,2}\gamma_{1,1},\\
&\varepsilon_{0,1}\gamma_{1,2} = \varepsilon_{1,2}\gamma_{0,0}.
\endaligned\tag3.3$$

And the solutions are  (with the $\varepsilon$'s given in the
 Section 4 of Ref. \cite{3}),
$$
\aligned
&\gamma (\varepsilon^I) = \pmatrix
 a & b & c\\
   & c & 0\\
   &   & 0\endpmatrix ,\
\gamma (\varepsilon^{II}) = \pmatrix
 a & b & c\\
   & 0 & 0\\
   &   & b\endpmatrix ,\\
&\gamma (\varepsilon^{III}) = \pmatrix
 a & b & c\\
   & 0 & 0\\
   &   & 0\endpmatrix ,\
\gamma (\varepsilon^{IV}) = \pmatrix
 a & 0 & 0\\
   & b & c\\
   &   & 0\endpmatrix ,\\
&\gamma (\varepsilon^{V}) = \pmatrix
 a & 0 & 0\\
   & 0 & b\\
   &   & c\endpmatrix ,\
\gamma (\varepsilon^{VI}) = \pmatrix
 a & 0 & 0\\
   & b & c\\
   &   & d\endpmatrix ,\\
&\gamma (\varepsilon^{VII}) = \pmatrix
 a & b & 0\\
   & 0 & 0\\
   &   & c\endpmatrix ,\
\gamma (\varepsilon^{VIII}) = \pmatrix
 a & 0 & b\\
   & c & 0\\
   &   & 0\endpmatrix ,\\
&\gamma (\varepsilon^{IX}) = \pmatrix
 0 & a & 0\\
   & b & c\\
   &   & a\endpmatrix ,\
\gamma (\varepsilon^{X}) = \pmatrix
 0 & 0 & a\\
   & a & b\\
   &   & c\endpmatrix ,\\
&\gamma (\varepsilon^{XI}) = \pmatrix
 0 & 0 & 0\\
   & a & b\\
   &   & c\endpmatrix ,\
\gamma (\varepsilon^{XII}) = \pmatrix
 a & b & 0\\
   & c & d\\
   &   & 0\endpmatrix ,\\
&\gamma (\varepsilon^{XIII}) = \pmatrix
 a & 0 & b\\
   & 0 & c\\
   &   & d\endpmatrix,\
\gamma (\varepsilon =(1)) = \pmatrix
 a & b & c\\
   & c & a\\
   &   & b\endpmatrix ,\endaligned
\tag3.4$$
where $a, b, c$ and $d$ are arbitrary.
 The matrix $\varepsilon =(1)$ has all its entries equal to 1 (it is
 not the identity matrix). Obviously, the $\gamma$'s corresponding to
 $\varepsilon =(0)$ are free.

\medskip

\leftline {\bf 3.3. $\Bbb Z_2\otimes\Bbb Z_2$-contractions.}
\medskip

The grading group consists of four elements, that we name
 $a=00,\ b=01,\ c=10$ and $d=11$, so that their product
 is $a+k=k,\ 2k=a\ (k=a,\dots ,d)$ and $b+c=d, c+d=b, b+d=c$.
 In the generic case ({\it i.e.} when all the $\varepsilon$'s are
 defined), the equations (2.12) are:
$$
\aligned
 &\varepsilon_{a,a}\gamma_{a,k} = \varepsilon_{a,k} \gamma_{a,k},\quad
 \varepsilon_{a,k}\gamma_{k,k} = \varepsilon_{k,k} \gamma_{a,a},\\
 &\varepsilon_{a,b}\gamma_{b,c} = \varepsilon_{a,c}\gamma_{b,c}
 = \varepsilon_{b,c}\gamma_{a,d},\\
 &\varepsilon_{a,b}\gamma_{b,d} = \varepsilon_{a,d}\gamma_{b,d}
 = \varepsilon_{b,d}\gamma_{a,c},\\
 &\varepsilon_{a,c}\gamma_{c,d} = \varepsilon_{a,d}\gamma_{c,d}
 = \varepsilon_{c,d}\gamma_{a,b},\\
 &\varepsilon_{b,c}\gamma_{d,d} = \varepsilon_{b,d}\gamma_{c,c}
 = \varepsilon_{c,d}\gamma_{b,b},\\
 &\varepsilon_{b,b}\gamma_{a,c} = \varepsilon_{b,c}\gamma_{b,d},\quad
 \varepsilon_{b,b}\gamma_{a,d} = \varepsilon_{b,d}\gamma_{b,c},\\
 &\varepsilon_{b,c}\gamma_{c,d} = \varepsilon_{c,c}\gamma_{a,b},\quad
 \varepsilon_{c,c}\gamma_{a,d} = \varepsilon_{c,d}\gamma_{b,c},\\
 &\varepsilon_{c,d}\gamma_{b,d} = \varepsilon_{d,d}\gamma_{a,c},\quad
 \varepsilon_{b,d}\gamma_{c,d} = \varepsilon_{d,d}\gamma_{a,b},
 \endaligned\tag3.5$$
where $k=b, c, d$.
The $\gamma$-solutions (corresponding to
 the $\varepsilon$-solutions given in Ref. \cite{3}) are in
 Tables I and II.

\medskip

\leftline {\bf 4. EXAMPLES.}
\medskip

\leftline {\bf 4.1. Contractions of Killing form.}
\medskip

First, consider the algebra $A_1$
 (the complexification of $sl(2)$), with elements $\{X, Y, Z\}$ and
 commutation relations $$[X,Y]=Z,\ [Y,Z]=X,\ [Z,X]=Y.$$ From these
 we find that the Killing form $\Omega^{Killing}(\cdot ,\cdot )
 \equiv Tr (ad(\cdot ) ad(\cdot ))$ is equal to
$$
\Omega^{Killing} = diag (-2, -2, -2).\tag4.1$$

Now consider the $\Bbb Z_2\otimes\Bbb Z_2$-grading
$$
L_b = \{X\},\ L_c = \{Y\},\ L_d = \{Z\},\tag4.2$$
where we use the same notation for the
 grading group elements as in subsection 3.3.
Following (2.8), the commutators are modified to
$$
[X,Y]_\varepsilon =\varepsilon_{b,c}Z,\
[Y,Z]_\varepsilon =\varepsilon_{c,d}X,\
[Z,X]_\varepsilon =\varepsilon_{b,d}Y,\tag4.3$$
so that the Killing form that we get from the adjoint
 representation is
$$
\Omega^{Killing,\varepsilon} =
 -2\ diag (\varepsilon_{b,c}\varepsilon_{b,d},\
\varepsilon_{b,c}\varepsilon_{c,d},\
\varepsilon_{b,d}\varepsilon_{c,d},).\tag4.4$$

Now if we apply the definition (2.10) for
 the grading (4.2), the form (4.1) becomes
$$
\Omega^{Killing, \gamma} = -2\ diag (\gamma_{b,b}, \gamma_{c,c},
 \gamma_{d,d}),\tag4.5$$
which is to be compared to (4.4). This is done by noting that the
 equations (2.12) which are relevant here ({\it i.e.} in which
 $\varepsilon$ and $\gamma$ do not
 contain $a$ as an index) are
$$
\varepsilon_{b,c}\gamma_{d,d} = \varepsilon_{c,d}\gamma_{b,b} =
 \varepsilon_{b,d}\gamma_{c,c},\tag4.6$$
for which
$$\gamma_{b,b}=\varepsilon_{b,c}\varepsilon_{b,d},\
 \gamma_{c,c}=\varepsilon_{b,c}\varepsilon_{c,d},\
 \gamma_{d,d}=\varepsilon_{b,d}\varepsilon_{c,d}\tag4.7$$
is a possible solution.
By comparing (4.5) to (4.4), we see that the form
 obtained by graded contractions of $A_1$ ({\it i.e.} by the
 introduction of $\varepsilon$-parameters into the commutators) can be
 compatible with the $\gamma$-contracted form obtained by the present
 procedure, {\it i.e.} using (2.10).

A similar compatibility exists in
 any general Lie algebra $g$ with basis $\{X_1,\dots ,X_{dim [g]}\}$ such
 that
$$[X_A, X_B] = f^C_{AB} X_C = ad(X_A)_{CB} X_C.\tag4.8$$
 Suppose that $X_A, X_B$ belong to
 the $\Gamma$-grading subspaces $g_\mu ,g_\nu$, respectively,
 so that $X_C\in g_{\mu +\nu}$. Then the Killing form is
$$
\aligned
\Omega^{Killing} (X_{\mu A}, X_{\nu B}) &=
 Tr \left [ad(X_{\mu A}) ad(X_{\nu B})\right ], \\
 &= \sum_{C,D} ad(X_{\mu A})_{(\mu +\sigma )C;\sigma D}\
      ad(X_{\nu B})_{(\mu +\nu +\sigma )D;(\mu +\sigma )C},
\endaligned\tag4.9$$
 from which we see that $\mu$ must be the $\Gamma$-inverse of $\nu$.
 (We have written the grading indices along with the algebra indices).

By introducing $\varepsilon$-parameters as in (2.8), this
 becomes
$$
 \Omega^{Killing, \varepsilon} (X_{\mu A}, X_{\nu B}) =
 \sum_{C,D} \varepsilon_{\mu ,\sigma}\varepsilon_{\nu ,\mu +\sigma}
 ad(X_{\mu A})_{(\mu +\sigma )C;\sigma D}
 ad(X_{\nu B})_{\sigma D;(\mu +\sigma )C},\tag4.10$$
where the $\varepsilon$'s are relevant if the corresponding commutators
 are defined. One can compare (4.10) to the form obtained via
 our definition (2.10), {\it i.e.}
$$
 \Omega^{Killing,\gamma} (X_{\mu A}, X_{\nu B}) =
 \gamma_{\mu ,\nu} \Omega^{Killing} (X_{\mu A}, X_{\nu B}),\tag4.11$$
where $\mu + \nu = 0$ (the identity element in $\Gamma$),
 if the algebra is be such that $\varepsilon^2$ in (4.10) can
 be factored out. Then one can identify
$$
\gamma_{\mu ,\nu} = \sum_\sigma \varepsilon_{\mu ,\sigma}
 \varepsilon_{\nu ,\mu +\sigma},\tag4.12$$
where the sum is over the $\sigma$
 indices such that the $\varepsilon$'s
 are defined, and with $\mu +\nu =0$.
In the particular case of $A_1$, (4.12) is just (4.7).

\medskip

\leftline {\bf 4.2. Toroidal contractions of $A_2$.}
\medskip

There are four different fine gradings of $A_2$
 (the complexification of $su(3)$) \cite{6--7}, one of
 which is the {\it toroidal} (or {\it Cartan}) grading:
$$
A_2 = \oplus_{\mu =-3}^3\ g_\mu,
$$
with
$$
\aligned
 g_1 &= \{e_\alpha\},\ g_{-1} = \{e_{-\alpha}\},\\
 g_2 &= \{e_\beta\},\ g_{-2} = \{e_{-\beta}\},\\
 g_3 &= \{e_{\alpha +\beta}\},\ g_{-3} =
 \{e_{-(\alpha +\beta )}\},\\
g_0 &= \{h_\alpha ,h_\beta\}.\endaligned\tag4.13$$
for which the grading group is $\Bbb Z_7$.

Consider the following invariant bilinear form on $A_2$;
$$
 \aligned
 \Omega (a_\delta ,a_{-\delta}) &= \Omega_\delta, \\
 \Omega (h_{\alpha_i}, h_{\alpha_j}) &= <\alpha_i |\alpha_j >,
 \endaligned\tag4.14$$
where $i, j = 1, 2$; $\delta$ is any root of $A_2$,
 and all other elements of $\Omega$ are zero.
 The only relevant $\gamma$-parameters are
 $\gamma_{0,0}$ and $\gamma_{\mu ,-\mu}$.

The grading (4.13) is such that the relevant
 $\varepsilon$-parameters are:
$$
\aligned
 &\varepsilon_{1,2},\ \varepsilon_{1,-3},\ \varepsilon_{1,-1},\
 \varepsilon_{-2,-1},\\
 &\varepsilon_{2,-3},\ \varepsilon_{2,-2},\ \varepsilon_{3,-3},\
 \varepsilon_{3,-2},\\
 &\varepsilon_{3,-1},\ \varepsilon_{0, \mu}\ \ (\mu = -3, \dots ,3).
 \endaligned$$

{}From (2.12), we get the following relations:
$$
\varepsilon_{\mu ,-\mu} \gamma_{0,0}
 = \varepsilon_{0,\mu} \gamma_{\mu ,-\mu}
 = \varepsilon_{0,-\mu} \gamma_{\mu ,-\mu},\tag4.15$$
for $\mu = 1, 2, 3$. The possible $\varepsilon$-parameters have been
 found and the associated algebras identified in Ref. \cite {8}.
 The $\gamma$ solutions are to be substituted into
$$
\aligned
&\Omega^\gamma (e_\alpha , e_{-\alpha}) = \gamma_{1,-1}\
 \Omega_\alpha,\\
&\Omega^\gamma (e_\beta ,e_{-\beta}) = \gamma_{2,-2}\
 \Omega_\beta,\\
&\Omega^\gamma (e_{\alpha +\beta}, e_{-(\alpha +\beta )}) =
 \gamma_{3,-3}\ \Omega_{\alpha +\beta},\\
&\Omega^\gamma (h_{\alpha_i}, h_{\alpha_j}) =
 \gamma_{0,0}\ <\alpha_i |\alpha_j>.\endaligned\tag4.16$$

The method can be applied readily to Kac-Moody algebras.  Some
 examples of graded contractions of Kac-Moody algebras are given
 in Ref. \cite {3}. (The particular case of Wigner-In\"on\"u
 contractions of Kac-Moody and Virasoro algebras has
 been studied explicitly in Refs \cite {9}
 and \cite{10}, respectively.)

\medskip

\leftline {\bf 4.3. Application to Wess-Zumino-Witten models.}
\medskip

In this subsection, we exploit the fact that,
 starting from a Lie algebra $g$
 with invariant bilinear form $\Omega$, one may define a WZW action
 on the corresponding group manifold \cite{11}. Suppose
 that the group $G$, which corresponds to
 the Lie algebra $g$, is simply connected.
 Consider a three-manifold $B$ with boundary $\Sigma = \partial B$
 and $\vartheta$, a map from $\Sigma$ to $G$, extended to a map from
 $B$ to $G$.
 The WZW action on a
 Riemann surface $\Sigma$ is \cite {11}
$$
S_{WZW} (\vartheta ) = {1\over{4\pi}}\int_\Sigma d^2 y\
 \Omega_{AB} A^A_aA^{Ba}\ +\ {i\over{12\pi}} \int_B d^3 y\
 \epsilon_{abc}A^{Aa}A^{Bb}A^{Cc}\Omega_{CD}f^D_{AB},
\tag4.17$$
where $a, b, c$ represent indices over $\Sigma$ and $B$,
and $\vartheta^{-1}
\partial_a\vartheta = A^A_aX_A$ (the $X$'s being the generators of the
 Lie algebra $g$).

Now if we consider a $\Gamma$-grading of $g$, such that $X_A\in
 g_{\Cal A}, X_B\in g_{\Cal B}$, etc. ($\Cal A, \Cal B, \dots$ denote
 the grading labels of the subspace to which $X_A, X_B, \dots$ belong,
 respectively)
 then we get a new WZW action:
$$
\aligned
S^{\varepsilon ,\gamma}_{WZW} (\vartheta ) &=
 {1\over {4\pi}}\int_\Sigma d^2y\ \gamma_{\Cal A,\Cal B} \Omega_{AB}
 A^A_aA^{Ba}\ + \\
 & \ \ + \ {i\over{12\pi}}\int_B d^3 y\ \epsilon_{abc}
 A^{Aa}A^{Bb}A^{Cc}\gamma_{\Cal C,\Cal D}\Omega_{CD}
 \varepsilon_{\Cal A,\Cal B} f^D_{AB}.\endaligned\tag4.18$$
 The invariant
 bilinear form $\Omega^\gamma$ must be non-degenerate, {\it i.e.}
 the inverse
 $\Omega^{\gamma ,AB}$ of $\Omega^\gamma_{AB}$
 must be defined such that
 $\Omega^{\gamma ,AB}\Omega^\gamma_{BC} = \delta^A_C$.
 The construction of a WZW model based on a non-semisimple group
 has been discussed recently in Refs \cite{12} and \cite{13},
 for ungauged and gauged models, respectively.

{}From that point of view, we see that the contraction
 procedure may provide a new geometry (in particular, a new
 spacetime), by starting from the one associated to the noncontracted
 WZW model, and by ``deforming'' it compatibly with the deformation
 of the algebra. The concept
 of contraction has not been mentioned explicitly in Refs \cite {12, 13},
 although the invariant bilinear form used therein was obtained from a
 contraction procedure \cite {14}.
 To our knowledge, contraction methods have been used explicitly
 for the first time in Refs \cite {15, 16}. The contraction used in
 \cite {15} leaves the contracted algebras
 with the particular $\Bbb Z_3$ structure,
$$
\varepsilon = \pmatrix
 1 & 1 & 1\\
   & 1 & 0\\
   &   & 0\endpmatrix ,$$
 which suggests a
 generalization along our point of view. A detailed discussion
 about this topic is postponed to a future publication.

 The deformation
 of a geometry into another one, using the concept of contraction,
 has been studied in Ref. \cite{17} (although, using a completely
 different approach). Group contractions, interpreted as
 quasi-catastrophical connections between different geometries,
 or topological fluctuations in spacetime,
 have been considered in Ref. \cite{18}, establishing a
 relation between different models of the universe.

\medskip

\leftline {\bf Acknowledgement}
\medskip

The author is indebted to Drs
 R. C. Myers and R. T. Sharp for reading the
 manuscript, and to the Natural Sciences and Engineering
 Research Council of Canada for financial support.

\newpage

\leftline {TABLE I. $\Bbb Z_2\otimes\Bbb Z_2$-contractions of
 invariant bilinear
 forms.}

The $\gamma$-solutions corresponding to the contractions
 of algebras given in Table 1 of Ref.
 \cite {3}, on which they must be superposed.

$$
\aligned
&\pmatrix
 a & b & c & d\\
   & a & d & c\\
   &   & a & b\\
   &   &   & a\endpmatrix
\pmatrix
 a & b & c & d\\
   & 0 & d & 0\\
   &   & a & b\\
   &   &   & 0\endpmatrix
\pmatrix
 0 & a & 0 & b\\
   & c & b & d\\
   &   & 0 & a\\
   &   &   & e\endpmatrix
\pmatrix
 a & 0 & b & 0\\
   & c & 0 & d\\
   &   & a & 0\\
   &   &   & e\endpmatrix\\
&\pmatrix
 a & b & c & d\\
   & a & d & c\\
   &   & 0 & 0\\
   &   &   & 0\endpmatrix
\pmatrix
 a & b & c & d\\
   & 0 & d & 0\\
   &   & 0 & 0\\
   &   &   & 0\endpmatrix
\pmatrix
 0 & a & 0 & b\\
   & c & b & d\\
   &   & 0 & 0\\
   &   &   & e\endpmatrix
\pmatrix
 a & 0 & b & 0\\
   & c & 0 & d\\
   &   & 0 & 0\\
   &   &   & e\endpmatrix \\
&\pmatrix
 0 & 0 & a & b\\
   & 0 & b & a\\
   &   & c & d\\
   &   &   & e\endpmatrix
\pmatrix
 0 & 0 & a & b\\
   & 0 & b & 0\\
   &   & c & d\\
   &   &   & e\endpmatrix
\pmatrix
 0 & 0 & 0 & a\\
   & b & c & d\\
   &   & e & f\\
   &   &   & g\endpmatrix
\pmatrix
 0 & 0 & a & 0\\
   & b & c & d\\
   &   & e & f\\
   &   &   & g\endpmatrix \\
&\pmatrix
 a & b & 0 & 0\\
   & a & 0 & 0\\
   &   & c & d\\
   &   &   & e\endpmatrix
\pmatrix
 a & b & 0 & 0\\
   & 0 & 0 & 0\\
   &   & c & d\\
   &   &   & e\endpmatrix
\pmatrix
 0 & a & 0 & 0\\
   & b & c & d\\
   &   & e & f\\
   &   &   & g\endpmatrix
\pmatrix
 a & 0 & 0 & 0\\
   & b & c & d\\
   &   & e & f\\
   &   &   & g\endpmatrix\endaligned$$

\newpage

\leftline {TABLE II. $\Bbb Z_2\otimes\Bbb Z_2$-contractions of
 invariant bilinear
 forms.}

The $\gamma$-solutions corresponding to the contractions
 of algebras given in Table 2 of Ref.
 \cite {3}, on which they must be superposed.

$$
\aligned
&\pmatrix
 a & b & c & 0\\
   & 0 & 0 & 0\\
   &   & 0 & 0\\
   &   &   & d\endpmatrix
\pmatrix
 a & 0 & 0 & b\\
   & c & d & 0\\
   &   & e & 0\\
   &   &   & 0\endpmatrix
\pmatrix
 a & b & 0 & c\\
   & 0 & 0 & 0\\
   &   & d & 0\\
   &   &   & 0\endpmatrix\\
&\pmatrix
 a & 0 & 0 & 0\\
   & b & c & 0\\
   &   & d & 0\\
   &   &   & 0\endpmatrix
\pmatrix
 0 & a & b & 0\\
   & c & d & b\\
   &   & e & 0\\
   &   &   & 0\endpmatrix
\pmatrix
 a & b & 0 & c\\
   & d & 0 & e\\
   &   & 0 & 0\\
   &   &   & f\endpmatrix\\
&\pmatrix
 0 & a & b & 0\\
   & c & d & 0\\
   &   & e & a\\
   &   &   & 0\endpmatrix
\pmatrix
 0 & a & b & 0\\
   & c & d & b\\
   &   & e & a\\
   &   &   & 0\endpmatrix
\pmatrix
 a & 0 & b & c\\
   & 0 & 0 & 0\\
   &   & d & e\\
   &   &   & f\endpmatrix\\
&\pmatrix
 0 & 0 & 0 & 0\\
   & a & b & c\\
   &   & d & e\\
   &   &   & f\endpmatrix
\pmatrix
 0 & a & 0 & b\\
   & c & 0 & d\\
   &   & 0 & a\\
   &   &   & e\endpmatrix
\pmatrix
 0 & 0 & 0 & 0\\
   & a & b & c\\
   &   & d & e\\
   &   &   & f\endpmatrix\endaligned
$$

last three columns:

$$
\aligned
&\pmatrix
 a & 0 & b & c\\
   & d & 0 & 0\\
   &   & 0 & 0\\
   &   &   & 0\endpmatrix
\pmatrix
 a & b & c & d\\
   & 0 & 0 & 0\\
   &   & 0 & 0\\
   &   &   & 0\endpmatrix
\pmatrix
 a & b & c & 0\\
   & d & e & 0\\
   &   & f & 0\\
   &   &   & 0\endpmatrix\\
&\pmatrix
 a & 0 & 0 & 0\\
   & b & 0 & c\\
   &   & 0 & 0\\
   &   &   & d\endpmatrix
\pmatrix
 a & b & c & d\\
   & 0 & 0 & c\\
   &   & 0 & 0\\
   &   &   & 0\endpmatrix
\pmatrix
 0 & 0 & 0 & 0\\
   & a & b & c\\
   &   & d & e\\
   &   &   & f\endpmatrix\\
&\pmatrix
 a & 0 & 0 & 0\\
   & 0 & 0 & 0\\
   &   & b & c\\
   &   &   & d\endpmatrix
\pmatrix
 a & b & c & d\\
   & 0 & 0 & 0\\
   &   & 0 & b\\
   &   &   & 0\endpmatrix
\pmatrix
 a & 0 & 0 & b\\
   & c & d & 0\\
   &   & e & 0\\
   &   &   & 0\endpmatrix\\
&\pmatrix
 0 & 0 & 0 & 0\\
   & a & b & c\\
   &   & d & e\\
   &   &   & f\endpmatrix
\pmatrix
 0 & 0 & a & b\\
   & 0 & 0 & a\\
   &   & c & d\\
   &   &   & e\endpmatrix
\pmatrix
 a & b & c & d\\
   & 0 & 0 & c\\
   &   & 0 & b\\
   &   &   & a\endpmatrix\endaligned
$$

\enddocument

\newpage

\Refs

\ref\no1\by E. In\"on\"u and E. P. Wigner
\jour Proc. Nat. Acad. Sci. US\vol 39\yr 1953
\pages 510--524\endref

\ref\no2\by R. Gilmore\paper Lie Groups, Lie Algebras and Some of Their
 Applications\jour (Chap. 10), J. Wiley \& Sons,\yr 1974\endref

\ref\no3\by M. de Montigny and J. Patera\jour J. Phys. A: Math. Gen.
 \vol 24\yr 1991\pages 525--547\endref

\ref\no4\by R. V. Moody and J. Patera\jour J. Phys. A: Math. Gen.
 \vol 24\yr 1991\pages 2227--2257\endref

\ref\no5\by A. M. Bincer and J. Patera\jour J. Phys. A: Math. Gen.
 \vol 26\yr 1993\pages 5621--5628\endref

\ref\no6\by J. Patera and H. Zassenhaus\jour Lin. Alg. Appl.
 \vol 112\yr 1989\pages 87--159\endref

\ref\no7\by J. Patera\jour J. Math. Phys.\vol 30\yr 1989
\pages 2756--2762\endref

\ref\no8\by M. Couture, J. Patera, R. T. Sharp, and P. Winternitz
 \jour J. Math. Phys.\vol 32\yr 1991\pages 2310--2318\endref

\ref\no9\by P. Majumdar\jour J. Math. Phys.\vol 34\yr 1993
 \pages 2059--2065\endref

\ref\no10\by S. Schrans\jour Class. Quant. Grav.\vol 10
 \yr 1993\pages L173--L181\endref

\ref\no11\by E. Witten\jour Comm. Math. Phys.\vol 92
 \yr 1984\pages 455--472\endref

\ref\no12\by C. R. Nappi and E. Witten\paper A WZW model
 based on a non-semi-simple group\jour IASSNS-HEP 93/61
  (1993), hep-th/9310112\endref

\ref\no13\by K. Sfetsos\paper Gauging a non-semi-simple
 WZW model\jour THU 93/30, hep-th/9311010
 \endref

\ref\no14\by D. Cangemi and R. Jackiw\jour Phys. Rev. Lett.
 \vol 69\yr 1992\pages 233--236\moreref\paper D. Cangemi
 \jour private communication\endref

\ref\no15\by D. I. Olive, E. Rabinovici, and A. Schwimmer
\paper A class of string backgrounds as a semiclassical limit
 of WZW models\jour SWA 93-94/15, WIS 93/1-CS, RI 93/69
 (1993), hep-th/9311081\endref

\ref\no16\by K. Sfetsos\paper Exact string backgrounds from WZW
 models based on non-semi-simple group\jour THU 93/31
 (1993), hep-th/9311093\moreref \paper Gauged WZW models and
 non-abelian duality\jour THU 94/01 (1994),
 hep-th/9402031\endref

\ref\no17\by F. J. Herranz, M. de Montigny, M. A. del Olmo,
 and M. Santander\paper Cayley-Klein algebras as graded
 contractions of SO(N+1)\jour to appear in:
 J. Phys. A: Math. Gen.; hep-th/9312126 \yr 1994\endref

\ref\no18\by R. Mignani\jour Lett. Nuov. Cim.\vol 33\yr 1978
 \pages 349--352\endref
\endRefs
\bye